\documentstyle[12pt,world_sci]{article}

\newcommand{\beqn}{\begin{eqnarray}}
\newcommand{\eeqn}{\end{eqnarray}}
\def\spose#1{\hbox to 0pt{#1\hss}}
\def\lsim{\mathrel{\spose{\lower 3pt\hbox{$\mathchar"218$}}
     \raise 2.0pt\hbox{$\mathchar"13C$}}}
\def\gsim{\mathrel{\spose{\lower 3pt\hbox{$\mathchar"218$}}
     \raise 2.0pt\hbox{$\mathchar"13E$}}}
\def\simpropto{\mathrel{\spose{\lower 3pt\hbox{$\mathchar"218$}}
     \raise 2.0pt\hbox{$\propto$}}}
\def\sixbar{\overline{6}}
\def\fifteenbar{\overline{15}}
\def\fortybar{\overline{40}}
\def\fortyfivebar{\overline{45}}
\def\fiftybar{\overline{50}}
\def\seventybar{\overline{70}}
\def\beq{\begin{equation}}
\def\eeq{\end{equation}}
\def\barr{\begin{array}}
\def\earr{\end{array}}
\def\and{\qquad {\rm and } \qquad}
\def\etal{ {\it et al.} }
\def\ie{ {\it i.e.} }

\def\ZPC#1#2#3{{\sl Z.~Phys.} {\bf C#1}, #2 (#3)}
\def\PTP#1#2#3{{\sl Prog. Theor. Phys.} {\bf #1}, #2 (#3)}
\def\PRL#1#2#3{{\sl Phys. Rev. Lett.} {\bf #1}, #2 (#3)}
\def\PRD#1#2#3{{\sl Phys. Rev.} {\bf D#1}, #2 (#3)}
\def\PLB#1#2#3{{\sl Phys. Lett.} {\bf B#1}, #2 (#3)}
\def\PREP#1#2#3{{\sl Phys. Rep.} {\bf #1}, #2 (#3)}
\def\NPB#1#2#3{{\sl Nucl. Phys.} {\bf B#1}, #2 (#3)}
\def\smgaugegroup{{\rm SU(3)_c \otimes SU(2)_L \otimes U(1)_Y}}
\def\tev{{\rm TeV }}
\def\gev{{\rm GeV }}

\def\mgut{M_{\rm GUT}}

\def\mz{m_{\rm z}}

\def\threebar{\overline{3}}
\def\fivebar{\overline{5}}
\def\tenbar{\overline{10}}
\def\fivteenbar{\overline{15}}
\def\thirtyfivebar{\overline{35}}
\def\fortybar{\overline{40}}
\def\fortyfivebar{\overline{45}}
\def\fiftybar{\overline{50}}
\def\seventybar{\overline{70}}
\def\xbar{\overline{x}}
\def\msbar{\overline{\rm MS}}
\def\tanb{\tan\beta}

\def\threetenths{{3\over 10}}
\def\half{{1\over 2}}

\catcode`\@=11
\long\def\@makefntext#1{
\protect\noindent \hbox to 3.2pt {\hskip-.9pt
$^{{\ninerm\@thefnmark}}$\hfil}#1\hfill}                

\def\@makefnmark{\hbox to 0pt{$^{\@thefnmark}$\hss}}  

\def\ps@myheadings{\let\@mkboth\@gobbletwo
\def\@oddhead{\hbox{}
\rightmark\hfil\ninerm\thepage}
\def\@oddfoot{}\def\@evenhead{\ninerm\thepage\hfil
\leftmark\hbox{}}\def\@evenfoot{}
\def\sectionmark##1{}\def\subsectionmark##1{}}

\renewcommand{\thefootnote}{\fnsymbol{footnote}}

\newcounter{sectionc}\newcounter{subsectionc}\newcounter{subsubsectionc}
\renewcommand{\section}[1] {\vspace*{0.6cm}\addtocounter{sectionc}{1}
\setcounter{subsectionc}{0}\setcounter{subsubsectionc}{0}\noindent
        {\normalsize\bf\thesectionc. #1}\par\vspace*{0.4cm}}
\renewcommand{\subsection}[1] {\vspace*{0.6cm}\addtocounter{subsectionc}{1}
        \setcounter{subsubsectionc}{0}\noindent
        {\normalsize\it\thesectionc.\thesubsectionc. #1}\par\vspace*{0.4cm}}
\renewcommand{\subsubsection}[1]
{\vspace*{0.6cm}\addtocounter{subsubsectionc}{1}
        \noindent
{\normalsize\rm\thesectionc.\thesubsectionc.\thesubsubsectionc.
        #1}\par\vspace*{0.4cm}}

\newcounter{appendixc}
\newcounter{subappendixc}[appendixc]
\newcounter{subsubappendixc}[subappendixc]

\renewcommand{\appendix}[1] {\vspace*{0.6cm}
        \refstepcounter{appendixc}
        \setcounter{figure}{0}
        \setcounter{table}{0}
        \setcounter{equation}{0}
        \renewcommand{\thefigure}{\Alph{appendixc}.\arabic{figure}}
        \renewcommand{\thetable}{\Alph{appendixc}.\arabic{table}}
        \renewcommand{\theappendixc}{\Alph{appendixc}}
        \renewcommand{\theequation}{\Alph{appendixc}.\arabic{equation}}
        \noindent{\bf Appendix \theappendixc #1}\par\vspace*{0.4cm}}

\def\abstracts#1{{

\centering{\begin{minipage}{12.2truecm}\footnotesize\baselineskip=12pt\noindent
        \centerline{\footnotesize ABSTRACT}\vspace*{0.3cm}
        \parindent=0pt #1
        \end{minipage}}\par}}

\renewenvironment{thebibliography}[1]
        {\begin{list}{\arabic{enumi}.}
        {\usecounter{enumi}\setlength{\parsep}{0pt}
\setlength{\leftmargin 1.25cm}{\rightmargin 0pt}
         \setlength{\itemsep}{0pt} \settowidth
        {\labelwidth}{#1.}\sloppy}}{\end{list}}

\newcounter{itemlistc}
\newcounter{romanlistc}
\newcounter{alphlistc}
\newcounter{arabiclistc}

\newcommand{\fcaption}[1]{
        \refstepcounter{figure}
        \setbox\@tempboxa = \hbox{\footnotesize Fig.~\thefigure. #1}
        \ifdim \wd\@tempboxa > 6in
           {\begin{center}
        \parbox{6in}{\footnotesize\baselineskip=12pt Fig.~\thefigure. #1}
            \end{center}}
        \else
             {\begin{center}
             {\footnotesize Fig.~\thefigure. #1}
              \end{center}}
        \fi}

\newcommand{\tcaption}[1]{
        \refstepcounter{table}
        \setbox\@tempboxa = \hbox{\footnotesize Table~\thetable. #1}
        \ifdim \wd\@tempboxa > 6in
           {\begin{center}
        \parbox{6in}{\footnotesize\baselineskip=12pt Table~\thetable. #1}
            \end{center}}
        \else
             {\begin{center}
             {\footnotesize Table~\thetable. #1}
              \end{center}}
        \fi}

\def\@citex[#1]#2{\if@filesw\immediate\write\@auxout
        {\string\citation{#2}}\fi
\def\@citea{}\@cite{\@for\@citeb:=#2\do
        {\@citea\def\@citea{,}\@ifundefined
        {b@\@citeb}{{\bf ?}\@warning
        {Citation `\@citeb' on page \thepage \space undefined}}
        {\csname b@\@citeb\endcsname}}}{#1}}

\newif\if@cghi
\def\cite{\@cghitrue\@ifnextchar [{\@tempswatrue
        \@citex}{\@tempswafalse\@citex[]}}
\def\citelow{\@cghifalse\@ifnextchar [{\@tempswatrue
        \@citex}{\@tempswafalse\@citex[]}}
\def\@cite#1#2{{$\null^{#1}$\if@tempswa\typeout
        {IJCGA warning: optional citation argument
        ignored: `#2'} \fi}}

 1
 1
 1

\font\ninerm=cmr9

\begin{document}
\begin{flushright}
arch-ive/9506296\\
MPI-PhT/95-51\\
June 1995
\end{flushright}

\centerline{\normalsize\bf GAUGE COUPLING UNIFICATION}
\baselineskip=16pt
\centerline{\normalsize\bf  IN GENERAL SUPERSYMMETRIC
MODELS\footnote{
to be published in the Preceedings of the Ringberg Workshop:
"Perspectives for electroweak interactions in $e^+e^-$ collisions",
February 5--8, 1995}}

\centerline{\footnotesize Ralf Hempfling}
\baselineskip=13pt
\centerline{\footnotesize\it Max-Planck-Institut f\"ur Physik,
Werner-Heisenberg-Institut,}
\baselineskip=12pt
\centerline{\footnotesize\it F\"ohringer Ring 6, 80805 Munich, Germany}
\centerline{\footnotesize E-mail: hempf@iws180.mppmu.mpg.de}
\vspace*{0.3cm}

\vspace*{0.9cm}
\abstracts{
We study the effects of additional fields
on the unification of gauge couplings
in supersymmetric models.
We find that the effects are quite constrained by the requirement
of SU(5) gauge invariance.
In general, we find
that any extension of the MSSM will
form complete SU(5) multiplets
or spoil gauge coupling unification.
}

\normalsize\baselineskip=15pt
\setcounter{footnote}{0}
\renewcommand{\thefootnote}{\alph{footnote}}
\section{Introduction}

The standard model of elementary particle physics (SM) is in
excellent agreement with present experimental results.
Nonetheless, the theory suffers from a variety of
theoretical shortcomings and is generally believed to
be the low energy effective part of a more fundamental
theory which is characterized by a higher symmetry
and fewer arbitrary parameter.
One very popular and promising step in this direction is to embed the
$\smgaugegroup$ in a single simple gauge group.
The most economical candidate for such a unified gauge group
is SU(5). Here, all fermions of one generation can be embedded in
a 5 dimensional and a 10 dimensional representation\cite{gutreview}.
The quantization of the U(1) hypercharges follows
automatically and the numerous constraints from
the absence of triangle anomalies reduces significantly.

Maybe the most prominent feature of such a grand unified
theory (GUT) is that baryon and lepton number are no
longer conserved separately as a consequence of having quarks and leptons
in the same multiplet.
As a result, the proton is no longer a stable particle.
Thus, the observed lower limit of the proton life-time
implies a severe lower bound on the scale of the GUT symmetry breaking,
$\mgut$.

Unfortunately, the SM as the complete low energy theory
of an SU(5) GUT model is ruled out because (a) the gauge
couplings do not meet within the theoretical and experimental errors,
(b) the large hierarchy between $\mgut$ and the electro-weak scale
is unstable under radiative corrections and would require excessive
fine-tuning
and (c) the predicted rate of proton decay is too fast.
However, all these problems can be avoided in the
minimal supersymmetric extension of the SM (MSSM).
Here, the prediction of the strong coupling constant,
$\alpha_s$, within the framework of the minimal supersymmetric model (MSSM)
obtained by assuming gauge coupling unification without any intermediate scale
is in acceptable agreement with
experiment\cite{amal}.
In addition, the unification of $\tau$ and bottom Yukawa couplings
is quite promising\cite{dhr}.

Despite these successes of supersymmetric grand unified theories (SUSY GUTs)
based on SU(5)\cite{su5} there remain still some problems.
Maybe the most sever challenge is the so-called
doublet/triplet problem\cite{doublet-triplet}
of giving the colored Higgs triplet a mass of the order of $\mgut$ while
retaining the Higgs doublets responsible for the electro-weak symmetry
breaking at the electroweak scale.
There have been many attempt to try and solve these problems
all of which have one thing in common:
they require the introduction of new fields. Thus, an
extension of the particle content of the MSSM is inevitable
and one might ask whether some of these new particles
can have a significant impact on the gauge coupling unification.
We will consider two cases. In section~2 we will
assume that the all the components of an SU(5) multiplet are present in
the low energy effective theory.
This idea has already been explored in refs.~\citenum{xsusy1,plb,xsusy2}.
Here any effect will come through multiplet splitting due
to renormalization group (RG) evolution.
In section~3 we will integrate the additional fields at $\mgut$.
Here the multiplet splitting  occurs through a possible coupling to the vacuum
expectation value (VEV) of the adjoint representation.

\section{Complete SU(5) Multiplets}

{}From the severity of the doublet/triplet problem we know
how hard it is to construct a model with a large mass hierarchy
between different SU(3)$\otimes$SU(2)$\otimes$U(1) members of
the same SU(5) multiplet. Therefore, we will first consider
extensions of the MSSM by complete SU(5) multiplets
in the low energy effective theory.

The renormalization group equations (RGEs) for these models
can be written at the one-loop level as
\beqn
{d \alpha_i\over d t} =  \alpha_i^2 \beta_i \,.
\label{rges}
\eeqn
Here, $t\equiv (2\pi)^{-1}\ln {(\rm scale)}$,
the indices $i,j = 1,2,3$ refer to the
U(1), SU(2) and SU(3) gauge group and summation over twice occurring
indices is assumed. Furthermore, the one-loop $\beta$ functions
for the gauge couplings are\cite{onelpgauge}
\beqn
\beta_i = \left(\matrix{0\cr-6\cr-9}\right)  +
  N_H    \left(\matrix{\threetenths\cr\half\cr0}\right)  +
\left(2 N_G + \beta^X\right)\,,
\label{rgeg}
\eeqn
where the three contributions to $\beta_i$ come from the gauge sector,
the Higgs doublets (in the MSSM the number of Higgs doublets, $N_H = 2$)
and the contribution of complete SU(5) multiplets.
It is a simple exercise to show that all such extensions
will preserve gauge coupling unification at the one-loop level.
The contributions of the gauge/gaugino sector to $\beta_i$ are
non-universal since some gauge bosons and their superpartners
acquire a mass via the Higgs mechanism
while others stay massless due to gauge invariance.
The contributions of the Higgs bosons are also non-universal because
the doublets are responsible for the electro-weak symmetry
breaking and should have a mass of the order of $\mz = 91.187~\gev$\cite{pdg}
while the Higgs triplets have to acquire a mass at $\mgut$
in order to sufficiently suppress the rate for proton-decay.
This doublet/triplet splitting\cite{doublet-triplet}
is solely motivated
by experiment and has no satisfying theoretical solution yet.
The last term which also contains the contribution of $N_G = 3$ generations
of quarks and leptons is universal for all three couplings.
(Note, that one family of quarks and leptons can be embedded in a
$\fivebar = d^c(\threebar,1,2/3)\oplus l(1,2,-1)$ and a
$10  = q(3,2,1/3)\oplus u^c(\threebar,1,-4/3)\oplus e^c(1,1,-2)$;
the numbers in brackets indicate the transformation properties under
the SU(3), SU(2) and U(1) gauge symmetries, respectively.)
The reason is that the inclusion of a full SU(5) multiplets with a
mass, $m$, does not break the SU(5) gauge symmetry and should yield a
universal contribution
to all three $\beta$ functions at any scale above $m$ at the one-loop level.
The contributions of the extensions of the MSSM can be written as
\beqn
\beta^X = \sum_\Phi T(\Phi)\,,
\label{betax}
\eeqn
where the sum is over all SU(5) multiplets $\Phi$.
The values of
$T(\Phi) \equiv d(\Phi) C_2(\Phi)/r$ are listed in Table~\ref{tphi}
for  all representations of SU(5) with $d<100$.
Here, $r = 24$ is the number of generators of SU(5) and $C_2(\Phi)$
[$d(\Phi)$]
is the quadratic Casimir operator [dimension]
of the SU(5) representation $\Phi$\cite{slansky}.
By imposing gauge coupling unification at $\mgut$ \ie
\beqn
\alpha_{\rm GUT}\equiv \alpha_1(\mgut)=\alpha_2(\mgut)=\alpha_3(\mgut)\,,
\label{bound}
\eeqn
and solving eq.~\ref{rges} to first order in perturbation theory we obtain
\beqn
t_0 \equiv {1\over 2\pi} \ln {\mgut\over \mz}
= {\alpha_1^{-1}(\mz) - \alpha_2^{-1}(\mz)\over \beta_1 - \beta_2} \simeq
5.3\,,\cr
\alpha_3^{-1}(\mz) = \alpha_2^{-1}(\mz) + t_0 (\beta_3 - \beta_2)  \simeq 8\,,
\label{solve}
\eeqn
where we have used
\beqn
\alpha_1(\mz) = {5\over 3} {\alpha_{\rm em}\over \cos^2 \theta_{\msbar}}\,,\cr
\alpha_2(\mz) =  {\alpha_{\rm em}\over \sin^2 \theta_{\msbar}}\,,\cr
\eeqn
and $\alpha^{-1}_{\rm em} = 127.9$
and $\sin^2 \theta_{\msbar} = 0.2319$\cite{pdg}
as the low energy input values.
The prediction of the strong coupling constant,
$\alpha_s(\mz) \equiv \alpha_3(\mz)
\simeq 0.125$ is in quite good agreement with the world average
$\alpha_s(\mz) = 0.117\pm0.005$~\cite{pdg}.
In deriving eq.~\ref{solve} we have assumed that there is
no intermediate scale but it also holds in the
case of a widely spread particle spectrum as long as the
members of the different SU(5) multiplets lie
close together.

\begin{table}[t]
\tcaption{The contribution of various SU(5) multiplets $\Phi$ to
$\beta^X$.}\label{tphi}
\small
$$
\begin{array}{|c||c|c|c|c|c|c|c|c|c|c|c|}
\hline
\Phi & 5,\fivebar & 10,\tenbar  & 15,\fivteenbar  & 24
&35,\thirtyfivebar& 40,\fortybar& 45,\fortyfivebar& 50,\fiftybar
&70,\seventybar& 70^\prime,\seventybar^\prime& 75
\\ \hline \hline
T(\Phi) & 1/2 & 3/2  &  7/2 & 5
& 14 & 11& 12 & 35/2 & 49/2 & 42 & 25
\\ \hline
\end{array}
$$
\end{table}

Note, that the right hand side of eq.~\ref{solve}
is independent of $\beta^X$.
This means that any extension of the MSSM by full SU(5)
multiplets will maintain the property of gauge coupling unification
at one loop.
The unification scale $\mgut \simeq 2 \times 10^{16}~\gev$
remains also unchanged and for the unified gauge coupling we obtain
\beqn
\alpha^{-1} \simeq 24 - t_0 \beta^X\,.
\label{limitd}
\eeqn
By requiring that the right hand side of eq.~\ref{limitd} is larger than
zero we find $\beta^X \lsim 4.5$ but maybe models with $\beta^X = 5$
are still
acceptable due to higher order corrections or threshold corrections
{\it etc.} and shall be included into our considerations.

In order to derive a viable model, we have to impose additional
constraints.
The cancellation of triangle anomalies and the fact
that all additional particles have to be massive
implies that complex representations
only occur in pairs.
Thus, there are four types of extensions satisfying the above requirements
\begin{itemize}
\item $n$ additional pairs of $5$ and $\fivebar$, where $n = 1,2,3,4,5$,
\item one additional pair of $10$ and $\tenbar$,
\item $n$ additional pairs of $\fivebar$ and $10$ where $n = 1,2$,
\item one additional adjoint representation,
$24 = g(8,1,0)\oplus w(1,3,0) \oplus b(1,1,0)
\oplus x(3,2,5/3) \oplus \xbar(\threebar,2,-5/3)$.
\end{itemize}
Experimental lower limits on the additional
particle masses can be satisfied by
adding explicit  dirac or majorana mass terms to the superpotential
allowed by gauge invariance in models of type 1, 2 and 4.
Models of type 3 correspond to the MSSM with four or five generations.
For the invisible width of the $Z$ boson at LEP experiments we
know that the number of (almost) massless neutrinos is 3 and a mechanism
has to be introduced in order to give mass to the additional
neutrinos larger than about $\mz/2$. Also, a lower limit on the
mass of an additional lepton of $m_{\tau^\prime}\gsim \mz/2$ at LEP
and a lower limit on the
mass of an additional down type quark of $m_{b^\prime}>85~\gev$ at CDF
have been established\cite{pdg}.
This implies that the Yukawa couplings for the additional
fermions are bound from below since no explicit gauge
invariant mass term exists.
On the other hand, there is an upper limit on the masses
from the infra-red fixed-point behavior of the Yukawa couplings
leaving only a very constrained region in parameter space.
The four generation model has been studied recently and found to be quite
constrained but could still be feasible if a right-handed
neutrino is introduced to raise the mass of the left-handed
neutrino above the experimental bounds\cite{gunion}.
The five generation model is even more
constrained and might already be ruled out by present data.
However, the model with additional $5, \fivebar, 10, \tenbar$ with the
possibility for a explicit dirac mass for all the
additional particles is still allowed.
It also does not require any additional fields
to generate mass for the unseen neutrinos.

\begin{table}[t]
\tcaption{The $\beta$ functions for the additional fields,
$\phi$.}\label{betaphi}
\small
$$
\begin{array}{|c||c|c|c|c|c|c|c|c|c|c|}
\hline
\beta^i_\phi&  l  & d^c  & e^c & u^c& q & b & w & g & x&\xbar
\\ \hline \hline
U(1) &-{3\over10}&-{2\over15}&-{6\over 5}&-{8\over15}&-{1\over30}
& 0 & 0 & 0 &-{5\over6} &-{5\over6}
\\ \hline
SU(2)&-{3\over 2}   & 0  & 0  & 0 &-{3\over 2} &
0 & -4 & 0 &-{3\over 2} &-{3\over 2}
\\ \hline
SU(3)& 0    &-{8\over 3}& 0 &-{8\over 3}&-{8\over 3}& 0& 0 &-6
&-{8\over 3}&-{8\over 3}
\\ \hline
\end{array}
$$
\end{table}

The assumption that all the members, $\phi$,
of one SU(5) multiplet, $\Phi$,  are mass
degenerate is protected by gauge invariance.
It acquires corrections below $\mgut$ where the gauge
symmetry is broken
through one-loop RG evolution
\beqn
{d \mu_{\phi}\over d t} = \mu_\phi\sum_i \alpha_i \beta_\phi^i\,,
\label{betamu}
\eeqn
where the $\beta^i_\phi$ are listed in Table.~\ref{betaphi}.
The splitting between the various members of an SU(5) at the electro-weak
scale
multiplet due to RG evolution is quite considerable and gives rise to
significant threshold corrections.
These corrections have to be included if we want to obtain a
self-consistent renormalization group result with two-loop $\beta$
functions.
If we decouple the fields $\phi$ from the RGEs at $\mu_\phi$
we obtain the improved one-loop formulas
\beqn
\alpha_i^{-1}(\mz) = \alpha_{\rm GUT}^{-1}+t_0 \beta_i
- \Delta^{\rm MSSM}_i
-{1\over 2\pi}\sum_\phi N_\phi \ln{\mu_\phi\over \mz} \beta_{i,\phi}\,,
\label{oneloop}
\eeqn
where $\beta_{i,\phi}$ are the contributions of the field, $\phi$, to
$\beta_i$ given in ref.~\citenum{massrge}. Furthermore,
\beqn
N_\phi = \left\{\matrix{N_5\cr N_{10} \cr N_{24}}\right\}~\hbox{for}
{}~\phi =   \left\{\matrix{l, d^c\cr e^c, u^c, q \cr b,w,g,x,\xbar}
\right.\,,
\eeqn
is the number of the fields $\phi$.
It has been shown in ref.~\citenum{antoniadis} that the step approximation
indeed contains the full one-loop threshold corrections
due to heavy particles.
The MSSM threshold corrections, $\Delta^{\rm MSSM}_i$,
studied in ref.~\citenum{mssmthresh} raise the predicted
value of $\alpha_s(\mz)$ by about 10\% and
spoil the success of the GUT prediction to some degree.

However, additional effects on the unification of the gauge couplings
arise because the universality of $\beta^X$ in eq.~\ref{rgeg}
is violated at the two-loop level.
The two-loop $\beta$ functions of these models derived
from ref~\citenum{jones} are listed in
ref.~\citenum{plb}.

In minimal SU(5) SUSY-GUTs the down-type quark fields $d$ and
the left-handed lepton fields $l$ are embedded in
one representation and as a result the
$\tau$ and bottom Yukawa couplings
are unified at $\mgut$.
By inspecting the $\beta$ functions it
is easy to see that $m_b/m_\tau$ increases with
$\alpha_{\rm GUT}$ and hence also with $\beta^X$.
This increase can be compensated by an increase in the top
Yukawa coupling, $\alpha_t$.
However, $\alpha_t$ quickly approaches its IR fixed-point
$\alpha_t = O(\alpha_s)$.
Hence, any significant effects of the top Yukawa coupling
can only come from the integration close to $\mgut$
and requires unperturbatively large values of $\alpha_t(\mgut)$.
Thus, in our numerical work we have used
$\alpha_t(\mgut) = \alpha_{\rm GUT}$ in order to obtain
a natural prediction for
$m_b/m_\tau$ for a
particular model assuming $\tau$-bottom Yukawa unification.

It is important to note that the additional particles discussed
here have an explicit SUSY conserving mass
parameter and are naturally heavier than the standard model particles
and their superpartners. Thus, by decoupling they will
be unconstrained by any presently available experimental data.

\begin{table}[t]
\tcaption{The one-loop (left column) and two-loop (right column) results for
the ratio, $r = \mgut^{X}/\mgut^{\rm MSSM}$,
$\Delta\alpha_s/\alpha_s$ [in \%],
$\alpha_{\rm GUT}$,
the ratio $m_b/m_\tau$.
We have chosen $m_0(\mgut) = 200~\gev$,
$\mu_\phi(\mgut) = 1~\tev$ for all additional fields, $\phi$,
and $\alpha_t(\mgut) = \alpha_{\rm GUT}$.
Furthermore, we have fixed $m_{1/2}$ such that
$M_{\tilde g}(M_{\tilde g}) = 200~\gev$.}\label{tresult}
\small
$$
\begin{array}{|c||c|c|c||c|c||c|c||c|c||c|c|}
\hline
{}&N_5 & N_{10} & N_{24}
    & \multicolumn{2}{c||}{r}
    & \multicolumn{2}{c||}{ \Delta \alpha_s/\alpha_s {\rm in \%} }
    & \multicolumn{2}{c||}{\alpha_{\rm GUT}}
    & \multicolumn{2}{c||}{ m_b/m_\tau }
\\ \hline \hline
{\rm MSSM}&
   0&0&0&  .72& 1.00&  -6.4&   0.0&   .040&   .042&  1.80&  1.87 
\\ \hline
1& 2&0&0&  .75& 1.11&  -4.8&   -.1&   .050&   .053&  1.88&  1.96 
\\ \hline
2& 4&0&0&  .79& 1.29&  -2.7&   -.1&   .066&   .073&  1.99&  2.09 
\\ \hline
3& 6&0&0&  .85& 1.59&   .43&   -.1&   .096&   .118&  2.15&  2.30 
\\ \hline
4& 8&0&0&  .97& 2.16&   6.3&  -1.8&   .175&   .333&  2.44&  2.65 
\\ \hline
5& 1&1&0&  .71& 1.63&  -6.1&   1.3&   .070&   .081&  2.08&  2.31 
\\ \hline
6& 3&1&0&  .76& 2.23&  -3.7&   1.5&   .104&   .149&  2.28&  2.64 
\\ \hline
7& 5&1&0&  .83& 4.20&   .73&  -2.1&   .203&   1.68&  2.64&  3.36 
\\ \hline
8& 0&2&0&  .48& 1.57&  -9.3&   -.2&   .088&   .118&  2.01&  2.28 
\\ \hline
9& 2&2&0&  .46& 2.16&  -7.6&  -1.6&   .144&   .332&  2.18&  2.62 
\\ \hline
10& 0&0&1&  .49&  .72&  -3.3&  26.&   .280&  2.38&  2.32&  3.26  
\\ \hline
\end{array}
$$
\end{table}

In Table~\ref{tresult} we have summarized our results for
the MSSM and 10 extended models characterized by
$N_5$, $N_{10}$, and $N_{24}$.
We have chosen the somewhat large values of
$\mu_\phi = 1~\tev$ for all additional fields $\phi$
in order to exploit as many models as possible.
In is clear that we recover the MSSM predictions if we raise
$\mu_\phi$ to $\mgut$.
The parameter independent lower limit on the gluino mass of
$M_{\tilde g} > 100~\gev$ has been established from direct
particle search at CDF\cite{refgluino}.
However, stronger limits can be
derived from the chargino/neutralino search by imposing
GUT constraints and we chose
$M_{\tilde g} = 200~\gev$ in order to safely
avoid all the present bounds.

The first row corresponds to the MSSM where we have
$\alpha_s^{\rm MSSM} = 0.124$ and
$\mgut^{\rm MSSM} = 2.3\times 10^{16}~\gev$
for our choice of parameters.
In the different columns we present the result for the
unification scale divided by the two-loop MSSM value and denoted by $r$,
the relative change in the prediction of $\alpha_s$
with respect to the two-loop MSSM value,
the unified gauge coupling, $\alpha_{\rm GUT}$, and the ratio
of $m_b(\mz)$ to $m_\tau(\mz)$.
The two values correspond to the results obtained by using one-loop
and two-loop $\beta$ functions.
We see that already for the MSSM the
value of $m_b/m_\tau$ for $\alpha_t(\mgut) = \alpha_{\rm GUT}$
is slightly above its experimental value of $m_b/m_\tau \simeq 1.6$
but can still be brought in agreement with
experiment by a modest increase of $\alpha_t$
or by choosing $\alpha_b = O(\alpha_t)$\cite{dhr}.
The situation becomes more problematic in all
extended models.
On the one hand, this ratio increases even more due to larger values of
$\alpha_s$ at high scales.
On the other hand, the IR fixed point of the top quark Yukawa coupling
obtained from
\beqn
{d\over dt}{\alpha_t\over \alpha_s} =
(12 - \beta^X){\alpha_t\over \alpha_s} -12 = 0 \,,
\eeqn
increases with $\beta^X$.
As a result, the resent measurement of the top quark mass,
$m_t = 176\pm8({\rm stat})\pm10{\rm (sys)}~\gev$
by the CDF collaboration\cite{refmt}
implies that either the top Yukawa coupling is below its
IR fixed point or one has to move even closer to the
theoretically unfavorable value $\tanb = 1$
(the solution for $\tanb \simeq m_t/m_b$
might be ruled out entirely for large enough
values of $\beta^X$).

Furthermore, we see that the models 4, 7, 9 and 10 become
non-perturbative at $\mgut$.
This scenario of non-perturbative unification
was already advocated in ref.~\citenum{nonpert1} in non-SUSY models
and in ref.~\citenum{nonpert2} extended to SUSY models
as being particularly attractive.
The reason is that the dependence of $\alpha_i$ on
$\alpha_{\rm GUT}$ in eq.~\ref{oneloop} vanishes in the
large $\alpha_{\rm GUT}$ limit.
It is interesting that the predictability of the models with
non-perturbative unification is not limited by large uncertainties
at $\mgut$ where the couplings are large but by
low energy threshold corrections.
This is due to the fact that the masses of the various members of one
SU(5) representation are widely scattered due to RG evolution
in eq.~\ref{betamu}.

\section{SU(5) Multiplets with Mass Splitting}

\begin{figure}
\vspace*{13pt}
\vspace*{8.truein}             
\includegraphics{m1.ps}
\includegraphics{m2.ps}
\includegraphics{m4.ps}
\includegraphics{m3.ps}
\fcaption{Threshold corrections to $\alpha_s$ and $\mgut$ due to
a pair of various complex SU(5) representations $\phi$ and $\bar \phi$
with
(a) $\phi = 5,10,15$
(b) $\phi = 35,40,45$
(c) $\phi = 50,70,70^\prime$
and due to
the real SU(5) multiplets,
$\phi = 24, 75$.}
\label{fig1}
\end{figure}

Sofar we have neglected the effects of superheavy multiplets
({\it i.e.} multiplets that acquire mass at $\mgut$).
This assumption is justified in the case where all components
of one multiplet are mass degenerate as they have to be if SU(5)
is unbroken. However, in general the fields will not only obtain
a mass from an explicit mass term but also from an interaction term
$W \supset \lambda \bar\phi 24 \phi$,
where the 24 dimensional representation acquires a non-zero VEV,
$a$.
In this case we obtain GUT threshold corrections
to the gauge couplings from eq.~\ref{oneloop}
with the masses given by
\beqn
\mu_{\phi,\bar \phi} = \left|\mu + Y_{\phi,\bar \phi}\lambda a\right|\,.
\eeqn

\begin{table}[t]
\tcaption{The changes in $\alpha_s$ and $\mgut$ for an
additional pair of complex fields
with dimension, $d$.}\label{tablex0}
\small
$$
\begin{array}{|c||c|c|c|c|}
\hline
d &  5  & 10  & 15 & 35
\\ \hline \hline
 \alpha_s/\alpha_s^0 & -1.0 & 10. & 5. & 4.5
\\ \hline
 \mgut/\mgut^0 & 0.97 & 2.2 & 2.0 & 10.
\\ \hline
\end{array}
$$
\end{table}

In fig.~\ref{fig1} we present the threshold corrections
to $\alpha_s$ and $\mgut$ due to various SU(5) multiplets
with dimension $d<100$. The corrections shown
in fig.~\ref{fig1}(a)--(c) are due to
a pair of complex representations $\phi$ and $\bar \phi$
($\phi = 5, 10, 15, 35, 40, 45, 70, 70^\prime$)
and the ones in fig.~\ref{fig1}(d) are due to
one real representation $\phi = 24, 75$.
They are obtained by varying
$0< a/\mu <\infty$.
All curves start at
$(1, 0)$ for $a = 0$
since all the threshold corrections vanish
in this case.
The corrections diverge logarithmically if one of the particle masses
vanish. The curves terminate for $\mu = 0$.
The corresponding end points are listed in
table~\ref{tablex0} for all the models where
unification is possible within the $\alpha_s$--$\mgut$ plane
under consideration.
(It looks as if these end points would lie in the middle
of the contours rather than at the end.
This is due to the fact that two branches of the same curves coincide.)
These regions of large corrections require typically a severe fine-tuning.
There are two types of exceptions.
First, in the cases of real representations
the neutral fields under the U(1) symmetry can be naturally
massless while the charge fields can have an arbitrarily large mass.
Second, the cases where the multiplet splitting is achieved via
a missing-partner mechanism\cite{missing partner}.
In table~\ref{tabmpartner} we present all cases where
parts of an SU(5) multiplet can acquire a mass from a term
\beqn
W \supset \phi_1 24 \bar \phi_2~\hbox{or}~  \phi_1 75 \bar \phi_2\,,
\label{missp}
\eeqn
with the the dimensions of the representations smaller than 100.
The fields obtaining mass from eq.~\ref{missp} are the ones common
to $\phi_1$ and $\phi_2$ (third column of table~\ref{tabmpartner}).
There are three different possibilities for the low energy particle spectrum:
\begin{itemize}
\item only the fields common in $\phi_1$ and $\phi_2$
are heavy
(ie. $\beta_i^X = T(\phi_1)+T(\phi_2) - 2\Delta \beta_i$)
\item there is an additional explicit mass term
$\mu \phi_1 \bar \phi_1$
(ie. $\beta_i^X = T(\phi_2)+T(\bar\phi_2) - 2\Delta \beta_i$)
\item there is an additional explicit mass term
$\mu \phi_2 \bar \phi_2$
(ie. $\beta_i^X = T(\phi_1)+T(\bar\phi_1) - 2\Delta \beta_i$)
\end{itemize}
with the $ \Delta \beta_i$ listed in table~\ref{tabmpartner}.
(Note, that $\phi_i$ and $\bar \phi_i$ are different fields.)
It is a simple exercise to check that none of these possible extension yield
a feasible model.

\begin{table}[t]
\tcaption{
The various members of a SU(5) multiplet which can acquire a mass via
the missing partner mechanism.}\label{tabmpartner}
\small
$$
\begin{array}{|c|c||c||c|c|c|}
\hline
\phi_1 & \bar\phi_2 & \hbox{Heavy fields} & \Delta\beta_1 &
\Delta\beta_2 &\Delta \beta_3
\\ \hline \hline
5 & \fortyfivebar & (1,2,1)\oplus(3,1,-2/3) & 1/2 & 1/2 & 1/2
\\ \hline
5 & \fiftybar & (3,1,-2/3) & 1/5 & 0 & 1/2
\\ \hline
10 & \fifteenbar & (3,2,1/3) & 1/10 & 3/2 & 1
\\ \hline
10 & \fortybar & (3,1,-4/3)\oplus(3,2,1/3) & 9/10 & 3/2 & 3/2
\\ \hline
15 & \fortybar & (3,2,1/3) & 1/10 & 3/2 & 1
\\ \hline
35 & \fortybar & (\threebar,3,-4/3)\oplus(\sixbar,2,1/3) & 26/10 & 9 & 13/2
\\ \hline
45 & \fiftybar & (3,1,-2/3)\oplus(\threebar,2,-7/3)\oplus (8,2,1) & 59/10 &
11/2 & 15/2
\\ \hline
45 & \seventybar & (1,2,1)\oplus(3,1,-2)\oplus (3,3,-2/3) & 11/10 & 5/2 & 2
\\ \hline
\end{array}
$$
\end{table}

We see from fig.~\ref{fig1} in general an increase
in $\mgut$ favored by the lower limits on the proton decay\cite{pdecay}
results in an unfavorable increase in $\alpha_s$ and vice versa.
The exceptions are the $35$ and $70'$ dimensional representations
for some choices of $a/\mu$. Here, the agreements
with experiment can be mildly improved.

\section{Summary}

In summary, we have investigated various extensions of the MSSM
with additional SU(5) multiplets. We distinguish the cases of
(a) additional fields at the weak scale where the
mass splitting is generated via RG evolution and
(b) additional fields at $\mgut$ where the
mass splitting arises from a coupling to the VEV of the 24.
We find that the predictions for $\alpha_s$ and $\mgut$ are
quite stable in all models of type (a)
with the exceptions of those models with non-perturbative unification.
Here, the predictions will be softened by the low energy
threshold corrections due to a widely scattered unpredictable
particle spectrum.
On the other hand, the models of type (b) can have a large effect
due to the missing partner mechanism or fine-tuning.
However, no candidate could be found that would significantly improve
the predictions of the MSSM.
In conclusion, we can say that additional fields in the
low energy effective theory form complete SU(5) multiplets
or spoil gauge coupling unification.

\section{Acknowledgements}
I would like to thank J. Louis and S. Pokorski for useful conversations
as well as the organizers of this workshop
for their hospitality and for
creating such a stimulating and inspiring atmosphere.


\section{References}

\begin{thebibliography}{99}

\bibitem{gutreview}
For a review see e.g.,
R. Mohapatra, \sl Unification and Supersymmetry,
\rm 2nd ed. (Springer, Berlin 1992);
G.G. Ross, \sl Grand Unified Theories
\rm (Addison-Wesley, Reading, MA, 1984).

\bibitem{amal}
U. Amaldi, W. de Boer and H. F\"urstenau,
 \PLB {260}{443}{1991};
J. Ellis, S. Kelley and D.V. Nanopoulos, \PLB {260}{131}{1991};
P. Langacker and M.X. Lou, \PRD {44}{817}{1992}.

\bibitem{dhr}
S. Dimopoulos, L.J. Hall and S. Raby,
 \PRL {68}{1984}{1992};
\PRD {45}{4192}{1992};
V. Barger, M.S. Berger and P. Ohmann, \PRD {47}{1093}{1993};
M. Carena, S. Pokorski and C.E.M. Wagner, \NPB {406}{59}{1993};
P. Langacker and N. Polonsky, \PRD{47}{1093}{1993};
W.A. Bardeen, M. Carena, S. Pokorski and C.E.M. Wagner,
 \PLB{320}{110}{1994}.

\bibitem{su5}
H. Georgi and S.L. Glashow, \PRL{32}{438}{1974}.

\bibitem{doublet-triplet}
H. Georgi and S.L. Glashow, \NPB{193}{150}{1981}.

\bibitem{xsusy1}
L. Ib\'a${\rm\tilde n}$ez, \PLB{126}{196}{1983};
(E) {\bf B130}, 463 (1983).

\bibitem{plb}
R. Hempfling, MPI-preprint MPI-PhT/95-08, \sl Phys. Lett. \bf B\rm ,
to be published.

\bibitem{xsusy2}
J.E. Bj\"orkman and D.R.T. Jones, \NPB{259}{533}{1985}.

\bibitem{onelpgauge} M.B. Einhorn and D.R.T. Jones, \NPB{196}{475}{1981}.

\bibitem{pdg} Particle Data Group, L. Montanet \etal, \PRD{50}{1173}{1994}.

\bibitem{slansky} For a review of group theory for
unified model building, see: R. Slansky, \PREP{79}{1}{1981}.

\bibitem{gunion} J.F. Gunion, D.W. McKay and H. Pois, \PLB{334}{339}{1994}.

\bibitem{massrge} K. Inoue, A. Kakuto, H. Komatsu and S. Takeshita,
\PTP{67}{1889}{1982};
J.P. Derendinger and C.A. Savoy, \NPB{253}{285}{1985};
N.K. Falck, \ZPC {30}{247}{1986}.

\bibitem{antoniadis} I. Antoniadis, C. Kounnas and K. Tamvakis,
\PLB{119}{377}{1982}.

\bibitem{mssmthresh} G.G. Ross and R.G. Roberts, \NPB{377}{571}{1992};
P. Langacker and N. Polonsky, \PRD{47}{4029}{1993};
M. Carena, S. Pokorski and C.E.M. Wagner, \NPB{}{}{1993};
P.H. Chankowski, Z. Pluciennik and S. Pokorski, \NPB{439}{23}{1995}.

\bibitem{jones} D.R.T. Jones, \PRD{25}{581}{1982}.

\bibitem{refgluino} F. Abe {\it etal}
   [CDF Collaboration], \PRL{69}{3439}{1992}.

\bibitem{nonpert1} L. Maiani, G. Parisi and R. Petronzio,
\NPB{136}{115}{1978}.

\bibitem{nonpert2} N. Cabibbo and G.R. Farrar,
\PLB{110}{107}{1982}.
%
%
%
%

\bibitem{refmt}
F. Abe \etal\ [CDF-collaboration], FERMILAB-PUB-95/022-E,
\sl Phys. Rev. Lett.\rm , to be published.

\bibitem{missing partner}
A. Masiero, D.V. Nanopoulos, K. Tamvakis and T. Yanagida,
\PLB{115}{380}{1982};
B. Grinstein, \NPB{206}{387}{1982}.

\bibitem{pdecay} J. Hisano, H. Murayama and T. Yanagida,
\NPB{402}{46}{1993} and references therein.


\end{thebibliography}
\end{document}